\documentclass[prl,twocolumn,preprintnumbers,amsmath,amssymb,superscriptaddress]{revtex4-2}
\usepackage{graphicx}
\usepackage{dcolumn}
\usepackage{bm}
\usepackage{color}
\usepackage{amsmath}
\usepackage{amssymb}
\usepackage{latexsym}
\usepackage{multirow}
\usepackage{xcolor}
\usepackage{hyperref}
\usepackage[capitalise]{cleveref}
\usepackage{textcomp}
\begin{document}
\title{Non-Gaussian diffusion near surfaces}
\author{Arthur Alexandre$^a$}
\affiliation{Univ. Bordeaux, CNRS, LOMA, UMR 5798, F-33400, Talence, France.}
\author{Maxime Lavaud$^a$}
\affiliation{Univ. Bordeaux, CNRS, LOMA, UMR 5798, F-33400, Talence, France.}
\author{Nicolas Fares}
\affiliation{Univ. Bordeaux, CNRS, LOMA, UMR 5798, F-33400, Talence, France.}
\affiliation{Department of Physics, Ecole Normale Supérieure de Lyon, 69364, Lyon, France.}
\author{Elodie Millan}
\affiliation{Univ. Bordeaux, CNRS, LOMA, UMR 5798, F-33400, Talence, France.}
\author{Yann Louyer}
\affiliation{Univ. Bordeaux, CNRS, LOMA, UMR 5798, F-33400, Talence, France.}
\author{Thomas Salez} \email{thomas.salez@cnrs.fr}
\affiliation{Univ. Bordeaux, CNRS, LOMA, UMR 5798, F-33400, Talence, France.}
\author{Yacine Amarouchene}\email{yacine.amarouchene@u-bordeaux.fr}
\affiliation{Univ. Bordeaux, CNRS, LOMA, UMR 5798, F-33400, Talence, France.}
\author{Thomas Gu\'erin}\email{thomas.guerin@u-bordeaux.fr}
\affiliation{Univ. Bordeaux, CNRS, LOMA, UMR 5798, F-33400, Talence, France.}
\author{David S. Dean} \email{david.dean@u-bordeaux.fr}
\affiliation{Univ. Bordeaux, CNRS, LOMA, UMR 5798, F-33400, Talence, France.}
\affiliation{Team MONC, INRIA Bordeaux Sud Ouest, CNRS UMR 5251, Bordeaux INP, Univ. Bordeaux, F-33400, Talence, France.}

\begin{abstract}
We study the diffusion of particles confined close to a single wall and in double-wall planar channel geometries where the  local diffusivities depend on the distance to the boundaries. Displacement parallel to the walls is Brownian as characterized by its  variance, but it is non-Gaussian having a non-zero fourth cumulant. Establishing a link with  Taylor dispersion, we calculate the fourth cumulant and the tails of the displacement distribution for general diffusivity tensors along with  potentials generated by either the  walls or externally, for instance  gravity.  Experimental and numerical studies of the motion of a colloid in the direction parallel to the wall give measured fourth cumulants  which are correctly predicted by our theory. Interestingly, contrary to models of Brownian-yet-non-Gaussian diffusion, the tails of the displacement distribution are shown to be Gaussian rather than exponential. All together, our results provide additional tests and constraints for the inference of  force maps and local transport properties near surfaces. 
\end{abstract}
\maketitle

\def\thefootnote{a}\footnotetext{These authors contributed equally to this work.}

The transport properties of colloidal particles in complex media can be very different from those observed in simple fluids. In the bulk of simple fluids, beyond molecular length and time scales, the motion of a colloidal particle satisfies two important properties: (i) its Mean Squared Displacement (MSD) increases linearly with time (diffusive behavior) and (ii) the  probability distribution functions (PDF) of position increments are Gaussian. In complex media, exhibiting dynamical and spatial heterogeneities, or in presence of flows or active forces, both properties (i) and (ii) are generally not satisfied. Examples range from non-Gaussian transport in hydrodynamic flows, with consequences for chemical delivery in microfluidic environments \cite{aminian2016boundaries}, to  experimental observations of {\em anomalous} diffusion in  complex fluids and biological media \cite{bressloff2013stochastic,hofling2013anomalous,ernst2012fractional,tolic2004anomalous,shen2017single}. 

Colloidal dynamics in a large class of complex media can be described as either {\em Fickian-yet-non-Gaussian}, {\em anomalous-yet-Brownian} or {\em Brownian-yet-Non-Gaussian Diffusion} (BNGD).\cite{wang2009anomalous,wang2012brownian,skaug2013intermittent,leptos2009dynamics,guan2014even,chakraborty2020disorder}. These terms all refer to processes  with linear-in-time MSDs but non-Gaussian PDFs, usually having  exponential tails.  A generic explanation for this phenomenon is the {\em diffusing-diffusivity} mechanism Ref.~\cite{chubynsky2014diffusing}. In this scenario, BNGD is generated by a fluctuating diffusion constant, arising for example from  fluctuations of the local density or of the  gyration radius of complex macromolecules \cite{yamamoto2021universal}. The diffusing-diffusivity mechanism has been further explored \cite{chechkin2017brownian,sposini2018first,lanoiselee2018diffusion,lanoiselee2018model,jain2016diffusion,jain2016diffusingsurvival,yin2021non,miotto2021length,hidalgo2020hitchhiker}, but in most studies the assumptions invoked for the dynamically evolving diffusivity are generic but rather phenomenological. To the best of our knowledge, with the exception of the Brownian motion of an ellipsoidal particle \cite{han2006brownian,munk2009effective,czajka2019effects}, there is currently no experimental realization of a system exhibiting \textit{fluctuating diffusivity} at all times where the local diffusivity is quantified both experimentally, numerically and theoretically over broad spatial and temporal ranges.

In this Letter, we study the non-Gaussian diffusive motion of a colloidal particle near a hard wall. 
Hydrodynamic interactions at walls strongly modify the stochastic motion of neighboring objects \cite{felderhof2005effect,elgeti2015physics,jeney2008anisotropic,huang2015effect,choudhury2017active,hertlein2008direct,helden2015direct}. The local diffusivity  parallel to the wall depends on the distance to the wall which itself fluctuates due to diffusion perpendicular to the wall, thus generating the diffusing-diffusivity mechanism. Note that a similar situation was previously considered in Ref.~\cite{matse2017test}, but that study mainly focused on the motion  perpendicular to the wall: in this case, the non-Gaussian behavior due to diffusing diffusivity can be observed only at small times, since at long times the presence of an interaction potential with the wall induces non-Gaussian displacements -- even for uniform diffusivity. Furthermore, the non-Gaussianity of the motion in the parallel direction could not be resolved in Ref~\cite{matse2017test}.
Here, we focus on the motion parallel to the wall which is a genuine realization of diffusing diffusivity at all times. Our theoretical analysis  identifies a formal link with Taylor dispersion. In particular, we provide a calculation, that we verify experimentally and numerically,  of the fourth cumulant of the displacement along the walls, which quantifies non-Gaussianity at all times. We also show that the tails of the  PDF are Gaussian rather than exponential.  

\begin{figure}[t!]
	\includegraphics{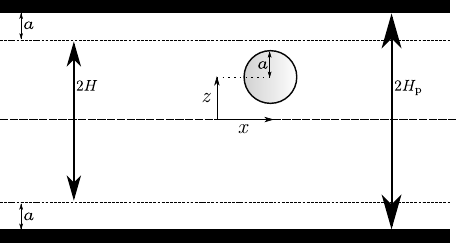}
	\caption{Schematic: a particle of radius $a$ diffuses in two dimensions, between two walls separated by a distance $2H_{\mathrm{p}}$.}
	\label{fig.shema}
\end{figure}

\begin{figure}[h!]
	\includegraphics[scale=1]{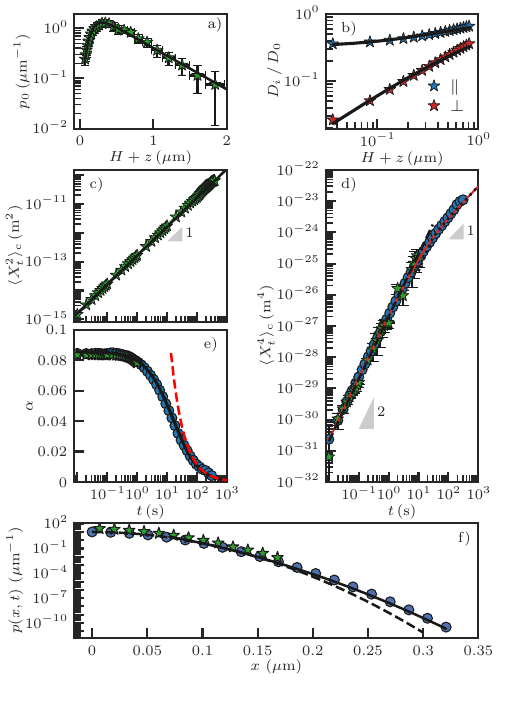}
	\caption{(a)~Experimental long-time PDF $p_0$ for position $z$, as a function of the distance $H+z$ to the bottom wall. Solid lines show the best fit to the Gibbs-Boltzmann distribution of Eq.~(\ref{gibbs_boltzmann}), using $V(z)$  in~\cref{Eq:PDF}, with $B= 5.0$, $l_\mathrm{D} = 88 ~\mathrm{nm}$ and $l_\mathrm{B} = 526 ~\mathrm{nm}$ at room temperature. Here, $H_{\textrm{p}}=40$~$\mu$m. Error bars give $95\%$ confidence intervals.  
	(b)~Experimentally-measured horizontal ($i=\parallel$) and vertical ($i=\perp$) local diffusion coefficients $D_i(z)$, normalized by the bulk value $D_0$. Solid lines are theoretical predictions $D_{i}(z) $ using~\cref{eq:d_perp,eq:d_para}.~(c)~Experimentally-measured second cumulant $\langle X_t^2 \rangle_{\textrm{c}}$. The solid line corresponds to~\cref{avd}, with $\langle D_{\parallel} \rangle_0 = 0.58 \, D_0$. The slope triangle indicates an exponent 1. (d)~Experimentally-measured (green stars) and numerically-simulated (blue dots) fourth cumulant $\langle X_t^4 \rangle_{\textrm{c}}$.   
	The red solid line represents the prediction of~\cref{c43}, while the black dashed and dotted lines are its short-time and long-time  asymptotics  of \cref{c4_short_time,def_c4}, with no adjustable parameter. The slope triangles indicate exponents 2 and 1. (e) Non-Gaussianity parameter $\alpha\equiv \langle X_t^4\rangle_{\text{c}}/\langle X_t^2\rangle_{\text{c}}^2$ as a function of time $t$, computed from the data and theoretical predictions in (c) and (d). The red dashed line shows the predicted  $\sim1/t$ decay (see SI~\cite{SM}). (f) PDF in horizontal displacement, at time increment $t=0.01~\mathrm{s}$, from experiments (green stars) and numerical simulations (blue dots). The Gaussian tail predicted in Eq.~(S45) (see SI~\cite{SM}) is also shown (solid line), as well as a simple Gaussian distribution (dashed line) with the proper mean and variance of the data.
}
	\label{fig.exp}
\end{figure}

\textit{Physical model. } We consider a Brownian particle of radius $a$ that is confined between two walls separated by a distance $2H_{\text{p}}$, as shown in Fig. \ref{fig.shema}. The particle diffuses along the channel (\textit{i.e.}~the $x$-axis) and perpendicularly to it  (\textit{i.e.}~the $z$-axis) with respective height-dependent diffusion coefficients $D_{\parallel}(z)$ and $ D_{\perp}(z)$, and is subject to a potential $V(z)$. The probability density   $p(z,t)$ about $z$ at time $t$ thus obeys the Fokker-Planck equation 
$\partial_t p = -\mathcal{H} p(z,t) $, where
\begin{equation}
\mathcal{H} \cdot = -\frac{\partial }{\partial z}\left\{D_\perp(z)\left[ \frac{\partial }{\partial z} \cdot +\beta \frac{\partial V(z)}{\partial z} \cdot \right]\right\}\ ,
\label{Hop}
\end{equation}
with $\beta = 1/(k_{\text{B}}T)$, where $k_{\text{B}}$ is Boltzmann's constant and $T$ is the temperature. We assume no-flux conditions at the walls. 
In the long-time limit, the system equilibrates along the $z$ direction and attains a Gibbs-Boltzmann distribution (see Fig. \ref{fig.exp}(a)):
\begin{equation}
	p_0(z) = \frac{e^{-\beta V(z)}}{ \int_{-H}^{H} e^{-\beta V(z')} dz'}\ . \label{gibbs_boltzmann}
\end{equation} 
We denote by $Z_t \in \left[-H, H \right]$   the height  of the center of the particle, $H = H_{\text{p}} -a$ the effective channel height available to the particle, and  $X_t$  the position of the center of the particle along the channel. 
The second and fourth cumulants
\begin{align}
&\langle X_t^2\rangle_{\text{c}}\equiv \langle X_t^2 \rangle, &
\langle X_t^4\rangle_{\text{c}}\equiv \langle X_t^4\rangle- 3\langle X_t^2\rangle^2\ ,
\end{align}
characterize the transport properties of the particle. Here, $\langle \cdot \rangle $ denotes the ensemble average, and the initial condition is $X_{t=0}=0$, while $Z_{t=0}$ follows the equilibrium distribution $p_0$. Note that $\langle X_t^4\rangle_{\text{c}}$  vanishes if $X_t$ is Gaussian, therefore its evaluation is the simplest way to quantify the non-Gaussian nature of the process $X_t$. 
We also define the related non-Gaussianity parameter $\alpha(t)\equiv \langle X_t^4\rangle_{\text{c}}/\langle X_t^2\rangle_{\text{c}}^2$. 

\textit{General theory. } The process $X_t$  obeys the stochastic differential equation:
\begin{equation}
    \label{eq:Xt}
	  dX_t = \sqrt{2D_\parallel(Z_t)} dB_{x,t}\ ,
\end{equation} 
where the Gaussian increments $dB_{x,t}$ have  $\langle dB_{x,t}\rangle =0$ and $\langle dB_{x,t}^2\rangle =dt$. 
In Eq. (\ref{eq:Xt}), we  use the Ito prescription of the stochastic calculus, however 
$D_\parallel(Z_t)$ is independent of $X_t$ and so this choice is unimportant.
Integrating Eq. (\ref{eq:Xt}) gives 
\begin{equation}
	  X_t =\int_0^t  \sqrt{2D_\parallel(Z_\tau)} dB_{x,\tau}\label{IntX}\ .
\end{equation} 
Squaring this and using the independence of $dB_{x,t}$ and $Z_t$, then taking the average  yields 
\begin{equation}
\langle X_t^2\rangle_{\text{c}}\equiv \langle X_t^2\rangle= 2 t \int_{-H}^{H} d z\,  D_\parallel(z) p_0(z) = 2\left< D_\parallel\right>_0 t\ ,\label{avd}
\end{equation}
where $\langle \cdot\rangle_0$ denotes the average with respect to the equilibrium distribution $p_0(z)$. The MSD is purely linear in time, so that   $X_t$ is Brownian at all times.  

Taking the fourth-power of  Eq.~(\ref{IntX}) and  using Wick's (or Isserlis') theorem \cite{coffey2012langevin}  gives~\cite{SM}
\begin{align}
\frac{\langle X_t^4\rangle_{\text{c}}}{12} =  \int_0^t ds \int_0^t ds' [ &\langle D_\parallel(Z_s) D_\parallel(Z_{s'})\rangle  - \langle D_\parallel  \rangle_0^2].
\label{kcumulant4}
\end{align}
Here, we  draw an analogy with  Taylor dispersion, for the dispersion of particles in channels in presence of hydrodynamic flows. We imagine the same process $Z_t$, but consider the convective displacement given along the channel by 
$Y_t=\int_0^t ds \ u(Z_s)$, where $u(z)$ is an arbitrary imposed flow field along the channel.
The first two  cumulants of $Y_t$ in this problem are 
\begin{align}
&\langle Y_t\rangle_\text{c} =\langle u \rangle_0 t,  \hspace{1cm}   
\nonumber\\ 
& \langle Y_t^2\rangle_\text{c}=  
 \int_0^t  ds \int_0^t ds' \ [  \langle u(Z_s) u(Z_{s'})\rangle  - \langle u  \rangle_0^2]. 
\end{align}
Comparing the above expressions with Eqs.~(\ref{avd}) and (\ref{kcumulant4}) we see that the second and fourth cumulants of $X_t$ are proportional to  the average and the variance of $Y_t$ in a Taylor dispersion problem with the formal correspondence $u(z)=D_\parallel(z)$.  
Taylor dispersion has been widely studied \cite{bar83,bis07,ved14,Li2015,gue15,gue15b,Vilquin2021,brennerMacrotransportProcesses1993a,mer94,bal95,mar19,watt1995accurate,alexandre2021generalized}, and we can exploit existing results for the MSD at all times from Ref.~\cite{alexandre2021generalized}, yielding the explicit expression: 
\begin{equation}
\begin{split}
\langle X_t^4\rangle_{\text{c}} &= 24 \int_{-H}^{H} dz \int_{-H}^{H}  dz'\, D_\parallel(z)D_\parallel(z') p_0(z')  \\
&\times  \sum_{\lambda>0}\left[\frac{t}{\lambda} -\frac{1-e^{-\lambda t}}{\lambda^2}   \right]\psi_{\text{R} \lambda}(z)\psi_{\text{L} \lambda}(z')\ ,
\end{split}
\label{c43}
\end{equation}
where $\psi_{\text{R} \lambda}(z)$ and $\psi_{\text{L} \lambda}(z)$ respectively denote the right and left eigenfunctions of $\cal H$, with eigenvalue $\lambda$, and the normalization $\int_{-H}^H dz \ \psi_{\text{L} \lambda}(z) \psi_{\text{R} \lambda}(z)=1$. In practice, this general expression can be evaluated by   numerically  computing the eigenfunctions after discretizing the operator $\mathcal{H}$. This formula  simplifies at short times into (see SI~\cite{SM}):  
\begin{equation}
\left< X_t^4\right>_{\text{c}} \underset{t\rightarrow 0}{\simeq} 12\,t^2\left[\left< D_\parallel^2\right>_0 - \left< D_\parallel\right>_0^2\right]\ .
\label{c4_short_time}
\end{equation}
We see that the initial non-Gaussianity parameter $\alpha (t=0) $ is finite and is  proportional to the variance of $D_\parallel(z)$ with respect to the equilibrium distribution, as in Ref.~\cite{chubynsky2014diffusing}. The late-time behavior is (see SI~\cite{SM})
\begin{equation}
\langle X_t^4\rangle_{\text{c}} \underset{t\rightarrow +\infty}{\simeq} 24 \left(D_4 t - C_4 \right)\ ,
\label{def_c4}
\end{equation}
where the explicit  expression of $C_4$ is given in the Supplementary Information (SI)~\cite{SM}, and where:
\begin{equation}
D_4 =\left< \frac{\left[J(z)e^{\beta V(z)}\right]^2}{D_\perp(z)}\right>_0\ ,
\label{D4_long}
\end{equation}
with:
\begin{equation}
J(z) = \int_{-H}^z dz' e^{-\beta V(z')}\left[D_\parallel(z')-\langle D_\parallel\rangle_0\right]\ .
\end{equation} 
From the above we see that the non-Gaussianity parameter satisfies $\alpha(t)\propto1/t$, for large $t$. 

\textit{Analytic strong confinement theory. } We consider  the case of a narrow  channel where we approximate the local diffusion coefficients  by the  quadratic expressions~\cite{lau07,avn20,SM}:
\begin{equation}
D_\perp(z)\approx D_{\perp 0}\left(1-\frac{z^2}{H^2}\right), \  
D_\parallel(z)\approx D_{|| 0}\left(1-\frac{z^2}{H_{\text{s}}^2}\right) ,
\end{equation}
where $H_{\text{s}}$ is a characteristic distance that can be considered as a slip-like length if $H_{\text{s}} > H$. The coefficients $D_{\perp 0}$ and $D_{|| 0}$ depend on the effective channel height $H$ and on the particle radius $a$. In principle, the no-slip boundary condition must impose that $D_\parallel(z)$ is zero at the walls, which would imply that $H_\text{s}=H$. However, numerically, $D_\parallel(z)$ is found to be quadratic near the channel center and decays rapidly to zero near the walls \cite{avn20,SM}.
Here,  Eq.  (\ref{c43}) can be evaluated by noting that the eigenfunctions are the  Legendre polynomials $P_n(z/H)$ of degree $n$, leading to:
 \begin{equation}
\begin{split}
\frac{\langle X_t^4\rangle_{\text{c}}}{24} & =   \frac{2 D_{||0}^2H^6 t }{135 D_{\perp 0}H_{\text{s}}^4} -\frac{D_{||0}^2H^8\left[1-e^{ - \frac{6 D_{\small\perp 0} t}{H^2}}\right]}{405 D^2_{\perp 0}H_{\text{s}}^4}.
\end{split}
\label{legendrePP}
\end{equation}  
In this model, there is only one relaxation time, equal to the time for the particle to diffuse perpendicularly.   This result can be generalized to arbitrary  $D_\parallel(z)$, keeping the same form of $D_\perp$, in which case many relaxation times appear (see SI~\cite{SM}). The non-Gaussianity is such that  $\alpha(0)= (12 H^4)/(5 (3 H_\text{s}^2-H^2 )^2)$ and is thus of order one. At short times,  if one takes $H_\text{s}=H$ then $\alpha(0)=3/5$. 
 
\textit{ Experimental system.} 
A polystyrene bead, with radius $a = 1.519 \pm 0.009~\mu\mathrm{ m}$ diffuses in an aqueous NaCl solution confined between two glass walls. Its trajectory  is tracked in three dimensions using Mie holography~\cite{lavaud2021stochastic}. Here, $H_\mathrm{p}=40$~$\mu$m, so that $H_\mathrm{p}\gg a$. The density mismatch of the particle is chosen such that the particle is visibly localized near the lower wall due to gravity (see~\cref{fig.exp}(a)). So, the effect of the upper wall is negligible both in terms of hydrodynamic and conservative forces. 
The bead is submitted to a potential:
\begin{equation}
\beta V(z)=\displaystyle B\,e^{-\frac{H+z}{l_\mathrm{D}}} +\displaystyle B\,e^{-\frac{H-z}{l_\mathrm{D}}}+ \frac{z}{l_\mathrm{B}} .
\label{Eq:PDF}
\end{equation}
The first two terms of the right-hand side are the screened electrostatic interactions 
between the   negatively-charged surfaces of the  walls and the bead, as given  mean-field theory~\cite{israelachvili2011intermolecular}, where $l_\mathrm{D}$ is the Debye length and $B$ is a dimensionless number depending in particular on the wall and bead surface charges. We have used the superposition approximation, valid for  gaps large compared to $l_\mathrm{D}$ so that the two potentials can be simply summed.
The third term accounts for gravity: $l_\mathrm{B}= k_\text{B}T/(\frac{4}{3}\pi a^3 \Delta\rho g)$  is the Boltzmann length, with $g$ the gravitational acceleration, and $\Delta\rho$ the density mismatch  between the polystyrene bead and the solution. 
Equations~\eqref{gibbs_boltzmann}~and~\eqref{Eq:PDF} are used to fit the experimentally measured  equilibrium distribution $p_0(z)$. The agreement is good with $B= 5.0 \pm 0.3$, $l_\mathrm{D} = 88 \pm 2 ~\mathrm{nm}$ and $l_\mathrm{B} = 526 \pm 5 ~\mathrm{nm}$, as shown in~\cref{fig.exp}(a). 
Assuming a perfect sphere, the value of $l_\mathrm{B}$ gives a density mismatch $\Delta\rho = 53 ~\mathrm{kg/m^3}$, which is within  5\% error  of the tabulated value of $50 ~\mathrm{kg/m^3}$.

Moving on to hydrodynamic interactions,  $D_\parallel$ and $D_\perp$ are inferred from the experimentally observed trajectories \cite{lavaud2021stochastic,SM} and are shown in~\cref{fig.exp}(b). The results agree with the Stokes-Einstein relations $D_i(z) = k_{\textrm{B}}T/[6 \pi a \mu_i(z)]$, where $i \in \{ \parallel, \perp \}$ 
and where $\mu_i$ are the components of the effective viscosity tensor~\cite{brennerSlowMotionSphere1961}. The transverse component reads~\cite{faxen1923bewegung}:
\begin{equation}
\mu_\parallel(z)= \mu_0\, \left(1 - \frac{9}{16}\zeta + \frac{1}{8}\zeta^3 - \frac{45}{256}\zeta^4 - \frac{1}{16}\zeta^5 \right)^{-1} , 
\label{eq:d_para}
\end{equation}
with $\zeta = a/(z+H_\text{p})$, and where $\mu_0$ is the bulk viscosity. The normal component was derived in Ref.~\cite{brennerSlowMotionSphere1961} as an infinite sum, which can be Pad\'e-approximated to within 1\% numerical accuracy by~\cite{bevanHinderedDiffusionColloidal2000} 
\begin{equation}
\mu_\perp(z)=  \mu_0 \,\frac{6\left(z+H\right)^2 +9a\left(z+H\right)+ 2a^2} {6\left(z+H\right)^2 +2a\left(z+H\right)}\ .
\label{eq:d_perp}
\end{equation}
These expressions, through the associated diffusion coefficients, are in agreement with the experimental data at room temperature and with $\mu_0=1$~mPa.s for water, as shown in~\cref{fig.exp}(b). Combined with the previously-mentioned equilibrium properties (see~\cref{Eq:PDF}), they can thus be used as inputs to compute  the theoretical values of the fourth cumulant of $X_t$.

\textit{Comparison with theory.}  Experimentally, the displacements $X_t$ are used to estimate $\langle X^2_t\rangle_\text{c}$ and $\langle X^4_t\rangle_\text{c}$, computed using the method of  sliding averages described in SI~\cite{SM} and leading to  Figs.~\ref{fig.exp}(c,d). First, the effective diffusion constant, $\langle  D_\parallel \rangle_0$ defined in~\cref{avd}, is given numerically by $\langle  D_
\parallel \rangle_0 = 0.58 \, D_0$, where  $D_0 = k_{\textrm{B}}T/(6 \pi a \mu_0)$ is the bulk diffusion constant. This is in  agreement with the experimental data shown in~\cref{fig.exp}(c). 
Secondly, the short-time theoretical prediction in~\cref{c4_short_time} correctly predicts the experimental data for $\langle X_t^4\rangle_\text{c}$, with no adjustable parameter (see~\cref{fig.exp}(d)). Lack of data at long times makes it difficult to check the late-time prediction given by~\cref{def_c4}. This result can however be verified through numerical simulations, as shown in~\cref{fig.exp}(d), where the simulation details are given in SI~\cite{SM}. 
The whole range of experimental and numerical data can be reproduced, up to error bars, by the exact prediction of~\cref{c43}, where the eigenfunctions and eigenvalues of $\mathcal{H}$ are computed numerically. We thus have a well-controlled experimental system with a MSD that is linear in time, at all times, as well as non-Gaussian statistics.

\textit{Distribution of displacements.} We now study the PDF $p(x,t)$ of the displacement $x$ at time $t$, in order to determine in particular whether or not it displays apparent exponential tails, as often observed in BNGD~\cite{wang2009anomalous,wang2012brownian,chakraborty2020disorder,chaudhuri2007universal,miotto2021length,rusciano2022fickian,xue2016probing,xue2020diffusion,pastore2021rapid,chechkin2017brownian}
 and other contexts \cite{silva2004exponential}. We consider a class of systems bounded in the $z$ direction, with a single maximum in $D_{||}(z)$, as is the case in our simulations, experiments and the simple channel model.
First, at short times, it is well known \cite{chubynsky2014diffusing,han2006brownian,chechkin2017brownian,lavaud2021stochastic} that 
\begin{align}
p(x,t)\simeq \int_{-H}^H dz \ p_0(z)\ \frac{e^{-x^2/{[4D_{\parallel}(z)t]}} }{\sqrt{4\pi D_\parallel(z) t}}\ .
\end{align} 
An analysis of this expression shows that, for large $x$, one has $p(x,t)\simeq A e^{-x^2/4D_{||}(z^*)t}$, where  $z^*$ is the point where $D_{||}(z) $ is maximal. 
In SI~\cite{SM}, we compute $A$, and we show in Fig.~\ref{fig.exp}(f) that such a Gaussian tail is quantitatively recovered in the numerical simulations and experiments.

At late times, the PDF can be analysed using its moment-generating function $g(q,t)=\langle e^{q X_t}\rangle$, which reads:
\begin{align}
g(q,t) = \left\langle e^{q\int_0^t dB_{x,s} \sqrt{2D_\parallel(Z_s)}}\right\rangle
=\left\langle e^{q^2\int_0^t ds  D_\parallel(Z_{s})}\right\rangle\label{gens}\ ,
\end{align}
where the last equality is obtained by averaging over the Gaussian noise $dB_{x,s}$. Interestingly, $g(q,t)=\langle e^{q^2Y_t}\rangle $ is related to the moment-generating function for the above-mentioned Taylor dispersion problem. We can thus use the tools introduced in the context of Taylor dispersion \cite{haynes2014dispersion,haynes2014dispersion1,Kahlen2017} to obtain the extreme tails of $p(x,t)$ at long times, with $p(x,t)\sim e^{-t f(x/t)}$. By extreme tails, we mean that $\xi = x/t$ is ${\cal O}(1)$, thus far from the diffusive scaling limit at late times where $x/\sqrt{t}$ is ${\cal O}(1)$. In SI~\cite{SM}, we show that for large $\xi$,
\begin{equation}
\small f(\xi)= \frac{1}{4D_{||}(z^*)}\left[\xi +{\rm sgn}(\xi)\sqrt{\frac{|D_{||}''(z^*)|D_\perp(z^*)}{2}}~\right]^2\ .\label{f_late_time}
\end{equation}
The paths contributing to $f(\xi)$ in this regime stay close to the region of maximal diffusivity, but they are rare and are exponentially suppressed with respect to paths which scale diffusively. 

The presence of Gaussian tails is thus generic for the class of problems studied here, and the exponential tails seen in other diffusing-diffusivity models \cite{chubynsky2014diffusing} are absent. 
The fact that $D_\parallel(z)$ has a maximal value is the  main difference between our system and those considered in  Ref.~\cite{chubynsky2014diffusing}, where the diffusion constant is unbounded. In fact, it was already noted in Ref.~\cite{chubynsky2014diffusing} that the PDF tails are generally not strictly exponential, depending on  the  local diffusivity distribution.  The Gaussian tails in our system contrast with the case of Continuous Time Random Walks  \cite{barkai2020packets}, and diffusion in confined disordered media \cite{chakraborty2020disorder,xue2016probing,xue2020diffusion} or glassy \cite{rusciano2022fickian,chaudhuri2007universal} systems, where exponentials tails are present. These situations tend to involve trapping, hoping or caging mechanisms (possibly due to heterogeneities) which are   absent in our system. 

The late-time corrections to Gaussianity in the diffusive-scaling region are dominated by the fourth-order cumulant, which gives a correction to the PDF that decays as $\sim1/t$  (see SI~\cite{SM}). Finally,  replacing $q=-ik$ in Eq.~(\ref{gens}) gives  the Fourier transform $\hat{p}(k,t)=\int_{-\infty}^\infty dx \ e^{-ik x}p(x,t)$. This can be computed  numerically (see SI~\cite{SM}). Taking the inverse Fourier transform gives a numerical evaluation of $p(x,t)$ in good agreement with the numerical and experimental PDFs (see Fig.~S2 in SI~\cite{SM}). We have also examined the case where the channel width is much smaller ($H_{\textrm{p}}=5.5\, \mathrm{\mu m}$). Similar effects are seen, but the asymptotic regime of linear temporal growth of the fourth cumulant is attained much more quickly. 

\textit{Conclusion}. We have addressed a  physical realization of diffusing-diffusivity motion based on confined colloids by establishing a mapping onto Taylor dispersion, where the diffusivity  formally corresponds to a flow field. This analogy gives quantitative predictions for the diffusion along the channel, which agree with experimental and numerical data with no additional fitting parameter apart from the physical ones obtained independently in the experiments. We have also shown that the tails of the PDF are not exponential but modified Gaussian in this generic class of models, both at short and long times. One should note that the effective diffusion constant along the channel only depends on the equilibrium properties of the process normal to the wall, and is otherwise independent of its dynamics. The fourth cumulant however depends on the precise details of the dynamics via the two-point probability density functions. The fourth cumulant thus carries extra information on the dynamics, and, as such, appears to be a key statistical observable that can further contribute to improve the experimental resolution for the inference of force maps and local transport properties in heterogeneous environments  \cite{serov2020statistical,frishman2020learning} and near surfaces \cite{lavaud2021stochastic}.

\begin{acknowledgments}
{\noindent \bf Acknowledgments:}The authors thank Joshua McGraw and Maxence Arutkin for interesting discussions. They acknowledge financial support from the European Union through the European Research Council under EMetBrown (ERC-CoG-101039103) grant. Views and opinions expressed are however those of the authors only and do not necessarily reflect those of the European Union or the European Research Council. Neither the European Union nor the granting authority can be held responsible for them. The authors also acknowledge financial support from the Agence Nationale de la Recherche under EMetBrown (ANR-21-ERCC-0010-01), Softer (ANR-21-CE06-0029), Fricolas (ANR-21-CE06-0039), and ComplexEncounters (ANR-21-CE30-0020) grants. Finally, they thank the Soft Matter Collaborative Research Unit, Frontier Research Center for Advanced Material and Life Science, Faculty of Advanced Life Science at Hokkaido University, Sapporo, Japan. 
\end{acknowledgments}

\end{document}


\title{Supplementary Information for \\ Non-Gaussian diffusion near surfaces}

\author{A. Alexandre,  M. Lavaud, N. Fares, E. Millan, Y. Louyer, T. Salez, Y. Amarouchene, T. Gu\'erin and D. S. Dean}
 
\bibliographystyle{naturemag}

\maketitle

\section{Theoretical formalism}
\subsection{Cumulants of the horizontal displacement}\label{cumulants}
The fourth cumulant is obtained by direct computation: 
\begin{equation}
	\langle X_t^4\rangle_{\text{c}}\equiv \langle X_t^4\rangle- 3\langle X_t^2\rangle^2 = 12 \int_0^t \dd s \int_0^t \dd s' \  \left[\langle D_\parallel(Z_s) D_\parallel(Z_{s'})\rangle  - \langle D_\parallel(Z_s) \rangle\langle D_\parallel(Z_{s'}) \rangle\right],
	\label{kcumulant4}
\end{equation}
where we have used  Wick's theorem. The translational invariance in the $x$ direction means that all odd cumulants are zero.
The cumulant can be rewritten as
\begin{equation}
	\begin{split}
 \langle X_t^4\rangle_{\text{c}}&= 12 \int_0^t \dd s \int_0^t \dd s' \left< \left[ D_\parallel(Z_s)-\left< D_\parallel(Z_{s})\right> \right] 
\left[D_\parallel(Z_{s'})-\left< D_\parallel(Z_{s'})\right> \right]\right>, \\
&= 12 \left< \left\{ \int_0^t \dd s \ \left[ D_\parallel(Z_s)-\left< D_\parallel(Z_{s}) \right> \right]\right\}^2\right>.
\end{split}
\label{refTaylor}
\end{equation}
The latter equation has the same Kubo-type structure as the second cumulant in the Taylor dispersion problem~\cite{alexandre2021generalized}. Interestingly, from the last expression in Eq.~(\ref{refTaylor}), we see that  $\langle X_t^4\rangle_{\text{c}}$ takes always positive values, regardless of the expression of $D_\parallel(z)$. 

To proceed, we introduce the propagator $p(z|z';t)$, \textit{i.e.} the probability to go from $z'$ at time zero to $z$ at time $t$, for the process $Z_t$. The propagator obeys:
\begin{equation}
\frac{\partial p(z|z';t)}{\partial t} = -{\cal H} \ p(z|z';t)\label{fpp},
\end{equation}
where the operator $\cal H$ acts on the variable $z$, and is given by Eq. (\ref{main.Hop}) of the main text, with the initial condition $p(z|z';0)=\delta(z-z')$. 
In this framework, for the process $X_t$, when $Z_t$ starts from equilibrium, one has:
\begin{equation}\langle X_t^4\rangle_{\text{c}}= 24  \int_0^t \dd s \int_0^s \dd s'\, \int_{-H}^H \dd z\,   D_\parallel(z)   \int_{-H}^H \dd z' D_\parallel(z') \left[ p(z|z';s-s')-p_0(z)\right]p_0(z') .
\label{c42}
\end{equation} 
We now introduce the left and right eigenfunctions, respectively $\psi_{\text{L}\lambda}$ and $\psi_{\text{R}\lambda}$, of $\mathcal H$ which obey:
\begin{align} 
&\mathcal{H}^\dagger \psi_{\text{L}\lambda}= \lambda \psi_{\text{L}\lambda},  
&\mathcal{H}\psi_{\text{R}\lambda}= \lambda \psi_{\text{R}\lambda},
 \end{align}
with $\lambda$ the associated eigenvalue and $\mathcal{H}^\dagger$ the adjoint operator of $\mathcal{H}$, which is  in general not self-adjoint. The solution of~Eq. (\ref{fpp}) for $p(z|z';t)$ then has the  decomposition:
\begin{equation}
p(z|z';t)= \sum_\lambda \psi_{\text{R}\lambda}(z)\psi_{\text{L}\lambda}(z')\exp(-\lambda t).
\end{equation}
The right eigenfunctions satisfy the no-flux boundary condition:
\begin{equation}
	\left\{D_\perp(z)\left[\frac{\mathrm{d} \psi_{\text{R}\lambda}}{\mathrm{d} z} + \beta V'(z)\psi_{\text{R}\lambda}(z)\right]\right\}_{z=\pm H}= 0,
\end{equation} and one can show~\cite{gar09} that the left eigenfunctions satisfy the Neumann condition: $\frac{\mathrm{d}}{\mathrm{d} z}\psi_{\text{L}\lambda}(z)|_{z=\pm H} = 0$. The eigenfunctions corresponding to $\lambda=0$ can be written as: $\psi_{\text{R}0 }(z)= p_0(z)$ and $\psi_{\text{L}0 }(z)= 1$, so that they respect the normalization condition $\int \mathrm{d}z\, \psi_{\text{R}0 }(z)\psi_{\text{L}0 }(z)=1$.
Using this representation of $p(z|z';t)$ in the Kubo formula of~Eq. (\ref{c42}), the fourth cumulant can be rewritten as:
\begin{equation}
\langle X_t^4\rangle_{\text{c}} = 24 \int_{-H}^{H} \dd z \int_{-H}^{H} \dd z'\, D_\parallel(z)D_\parallel(z') p_0(z')   \sum_{\lambda>0}\left[\frac{t}{\lambda} -\frac{1}{\lambda^2} + \frac{\exp(-\lambda t)}{\lambda^2}  \right]\psi_{\text{R} \lambda}(z)\psi_{\text{L} \lambda}(z'),
\label{c43}
\end{equation}
which is Eq. (\ref{main.c43}) of the main text. In principle, Eq.~(\ref{c43}) can be computed explicitly if the relevant eigenfunctions and eigenvalues are known, however in most cases, they are not known explicitly. Nevertheless, they can still be computed numerically using standard numerical packages and thus used to predict the full temporal behavior of $\langle X_t^4\rangle_{\text{c}}$. In contrast, the short-time and long-time behaviors can be extracted analytically from~Eq. (\ref{c43}), as explained in the following subsection.

\subsection{Asymptotic behavior of the fourth cumulant}
In the limit where $t\to0$,~ Eq. (\ref{c43}) simplifies to:
\begin{equation}
\langle X_t^4\rangle_{\text{c}} \underset{t\rightarrow 0}{\simeq} 12\,t^2 \int_{-H}^H \dd z  \int_{-H}^H \dd z' D_\parallel(z)D_\parallel(z') p_0(z')\sum_{\lambda>0}\psi_{\text{R} \lambda}(z)\psi_{\text{L} \lambda}(z').
\label{_c4_short_time}
\end{equation}
Furthermore, the completeness relation leads to: 
\begin{equation}
\sum_{\lambda>0}\psi_{\text{R} \lambda}(z)\psi_{\text{L} \lambda}(z')= \delta(z-z') - p_0(z)\label{crel}.
\end{equation}
Then, Eq.~(\ref{_c4_short_time}) becomes:
\begin{equation}
\left< X_t^4\right>_{\text{c}} \underset{t\rightarrow 0}{\simeq} 12\,t^2\left[\left< D_\parallel^2\right>_0 - \left< D_\parallel\right>_0^2\right].
\label{c4_short_time}
\end{equation}
The short-time behavior of the fourth cumulant is thus quadratic in time, and is proportional to the variance of $D_\parallel$ with respect to the equilibrium measure.

The fourth cumulant can also be computed in the limit where $t\to\infty$, \textit{i.e.} for $t \gg \lambda_1^{-1}$, where $\lambda_1$ is the first non-zero eigenvalue of $\mathcal{H}$. This is done by using a formulation in terms of Green's functions~\cite{gue15,gue15b}, which can be shown to be intimately linked to the macro-transport theory~\cite{brennerMacrotransportProcesses1993a}. In this case,~Eq. (\ref{c43}) simplifies to:
\begin{equation}
\langle X_t^4\rangle_{\text{c}} \underset{t\rightarrow +\infty}{\simeq} 24 \left(D_4 t - C_4 \right),
\label{def_c4}
\end{equation}
with
\begin{equation}
D_4 =\int_{-H}^{H} \dd z\,\int_{-H}^{H} \dd  z' D_\parallel(z)D_\parallel(z') p_0(z') 
\sum_{\lambda>0}\frac{\psi_{\text{R} \lambda}(z)\psi_{\text{L} \lambda}(z')}{\lambda},
\label{d4_theo}
\end{equation}
and
\begin{equation}
C_4=  \int_{-H}^{H} \dd z\,\int_{-H}^{H} \dd z' D_\parallel(z)D_\parallel(z') p_0(z')\sum_{\lambda>0}\frac{\psi_{\text{R} \lambda}(z)\psi_{\text{L} \lambda}(z')}{\lambda^2}.
\label{c4_theo}
\end{equation}
Using the method described in~\cite{alexandre2021generalized} (see Section III. A therein), we obtain:
\begin{align}
&D_4 =\left< \frac{\left[J(z) e^{\beta V(z)}\right]^2}{D_\perp(z)}\right>_0 ,&
\label{D4_long}
J(z) = \int_{-H}^z \dd  z' \exp[-\beta V(z')][D_\parallel(z')-\langle D_\parallel\rangle_0].
\end{align}
This form is particularly useful to carry out numerical computations with arbitrary potentials and diffusion tensors. One can also show that:
\begin{align}
&C_4 =  \left< R^2\right>_0 - \left< R\right>_0^2, &R(z)= \int_{-H}^z \dd z'\ \frac{J(z')\exp\left[\beta V(z')\right]}{D_{\perp}(z')}.
\label{C4_long}
\end{align}
 
\subsection{Analytical solutions for narrow channels}
\begin{figure}[th!]
	\includegraphics[width=16cm]{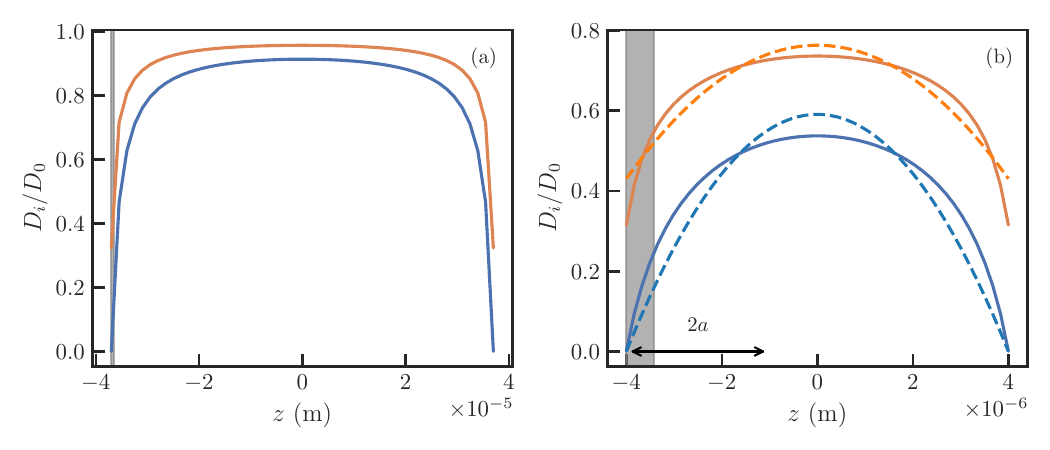}
	\caption{(a) Perpendicular and parallel diffusion coefficients, $D_\perp$ (blue) and $D_{||}$ (orange), as functions of vertical coordinate $z$, for a channel width $H_{\textrm{p}}=H+a=40$~$\mathrm{\mu m}$, in the superposition approximation (see Eq.~\eqref{eq:langevinDescretised}), using the parameters of the experiment. The grey region has width $l_{\textrm{B}}$ and represents the region where the particle remains typically localized due to gravity. (b) Same as previous panel, apart from the value of the width, which is  $H_{\textrm{p}}=H+a=5.5$~$\mathrm{\mu m}$. We also show here the quadratic approximations to $D_\perp$ and $D_{||}$ (dashed orange and blue lines). The arrow represents the diameter $2a$ of the bead.} 
	\label{figS1}
\end{figure}

In this part, we consider the simple case where there is no potential and where the channel has a sufficiently narrow width with respect to the particle size so that the diffusion constant can be taken to vary quadratically within the channel:
\begin{equation}
		D_\perp(z)= D_{\perp 0}\left(1-\frac{z^2}{H^2}\right), ~
		D_{||}(z)= D_{|| 0}\left(1-\frac{z^2}{H_{\text{s}}^2}\right) ,
	\label{dperp}
\end{equation}
where $H_{\text{s}}$ a characteristic length that can be considered as a diffusive slip length when $H_{\text{s}} > H$ and $D_\perp$ vanishes at $z = \pm H$. The coefficients $D_{\perp 0}$  and $D_{|| 0}$ depend on the effective channel height $H$ and $a$. The quadratic model of the diffusion constant in the height direction has been proposed in a theoretical context by a number of authors \cite{lau07,avn20}. In Fig.~\ref{figS1}(b), we have compared the superposition approximation for the local components of the diffusion tensor to simple quadratic fits for a narrow channel. As one might expect, due to the narrowness of the channel, this approximation works fairly well.

In the absence of an external potential, one has $p_0(z)=(2H)^{-1}$ and the effective longitudinal diffusion constant is given by:
\begin{equation}
	\left<D_{||} \right>_0= D_{|| 0}\left(1-\frac{H^2}{3H_{\text{s}}^2}\right) .\label{avds}
\end{equation}
Using Eqs.~(\ref{D4_long},\ref{C4_long}), we find:
\begin{equation}
	D_4 =\frac{ 2D^2_{||0} H^6}{135 D_{\perp 0}H_{\text{s}}^4} ,~
	C_4 =\frac{ D^2_{||0} H^8}{405 D_{\perp 0}^2H_{\text{s}}^4} .\label{dc4s}
\end{equation}
Since the diffusivities show quadratic profiles, a more detailed analysis, involving full time dependence is available. In fact, here the operator $\mathcal{H}$ (see Eq.~(\ref{main.Hop})) is self adjoint and its eigenvalues and normalized eigenfunctions  are given by:
\begin{align}
&	\psi_n(z) = \sqrt{\frac{2n+1}{2H}}\ P_n\left(\frac{z}{H}\right),&  \lambda_n = \frac{D_{||0}}{H^2}n(n+1),
\end{align} 
where $P_n$ denotes the $n^{th}$ degree Legendre polynomial.
 
 If we write $D_{||}(z)$ in terms of Legendre polynomials, we get:
\begin{equation}
	\frac{D_{||}(z)}{D_{||0}} =  \left(1-\frac{H^2}{3 H_{\text{s}}^2} \right) P_0\left(\frac{z}{H}\right)-\frac{2H^2}{3 H_{\text{s}}^2} P_2\left(\frac{z}{H}\right).
\end{equation}
From Eq. (\ref{c43}), the full time dependent behavior of the fourth cumulant is then given by:
\begin{equation}
	\begin{split}
		\frac{\langle X_t^4\rangle_{\text{c}}}{24} &=   \frac{2 D_{||0}^2H^6}{135 D_{\perp 0}H_{\text{s}}^4}t  -\frac{D_{||0}^2H^8}{405 D^2_{\perp 0}H_{\text{s}}^4} \left[1-\exp\left(-\frac{6D_{\perp 0} t}{H^2}\right)\right],
	\end{split}
	\label{SM_legendrePP}
\end{equation}
which is Eq. (\ref{main.legendrePP}) in the main text. From the latter, one can recover the late time corrections given in Eq. (\ref{dc4s}). 
This solution of the fourth cumulant can be extended to arbitrary expressions of $D_{\parallel}$ as long as it can be expressed on the basis of Legendre polynomials:
\begin{align}
&	D_{||}(z) = \sum_{n=0}^\infty d_n P_n\left(\frac{z}{H}\right) ,&
	d_n = \frac{2n+1}{2} \int_{-1}^1 \dd \zeta\ P_n(\zeta)D_{||}(\zeta H) .
\end{align}
The fourth cumulant reads in this general case:
\begin{equation}
	\begin{split}
		\frac{\langle X_t^4\rangle_{\text{c}}}{24} = &\sum_{n\geq 1}\left[\frac{H^2}{D_{\perp 0}n(n+1)}t -\frac{H^4}{D^2_{\perp 0}n^2(n+1)^2}  + \frac{H^4e^{- \frac{D_{\perp 0}}{H^2}n(n+1)t}}{D^2_{\perp 0} n^2(n+1)^2}\right]\frac{d_n^2}{2n+1} . 
	\end{split}
\end{equation}

\subsection{Probability distribution function of lateral displacement}
The probability distribution function (PDF) of the displacement $X_t$ can be analysed by considering the corresponding cumulant generating function:
\begin{equation}
g(q,t ) =\langle e^{q X_t}\rangle = \left\langle e^{q\int_0^t \sqrt{2 D_{||}(Z_s)} dB_{x,s}}\right\rangle   = \left\langle e^{q^2\int_0^t  D_{||}(Z_s) ds}\right\rangle,\label{gt1}
\end{equation}
where in the last equality we have averaged over the Brownian increments $dB_{x,s}$ which are independent of $Z_s$. 
 This has the form of a functional of the process $Z_s$: 
\begin{equation}
g(\mu,t) = \left\langle e^{\mu\int_0^t u(Z_s) ds}\right\rangle,
\end{equation}
where $\mu =q^2$ and $u(z)= D_{||}(z)$. Written this way, we see the mathematical resemblance between the diffusing diffusivity problem and Taylor dispersion in a velocity field $u(z)$:
\begin{equation}
Y_t = \int_0^t u(Z_s) ds,
\end{equation}
at the level of the generating functions of the two processes. The cumulant generating function, by definition, yields the cumulants of $Y_t$  via
\begin{equation}
g(\mu,t) = \exp \left[\sum_{n=1}^\infty \frac{\mu^n}{n!}\langle Y_t^n\rangle_{\text{c}} \right] .\label{843112}
\end{equation}
The functional $g(\mu,t)$  can  be evaluated by using the Feynman-Kac formula \cite{oksendal2003stochastic} as $g(\mu,t)=\langle G(\mu,t,z)\rangle_0$, where $G$ satisfies the equation:
\begin{align}
&\frac{\partial G(\mu,t,z)}{\partial t}= [-{\cal H}^\dagger + \mu D_\parallel(z)]G(\mu,t,z), & G(\mu,t=0,z)=1. 
\end{align}
Similar equations  appear for the diffusion of anisotropic objects~\cite{munk2009effective,kurzthaler2016intermediate}. 
If one determines the eigenvalues $\lambda(\mu)$ of $ {\cal H}^\dagger - \mu D_\parallel(z)$, then we obtain:
\begin{align}
g(\mu,t)=\sum_{\lambda(\mu)} e^{-\lambda(\mu)t} \  \int_{-H}^H dz \ p_0(z)\phi_{\text{R},\lambda }(z) \ \int_{-H}^H dz'\   \phi_{\text{L},\lambda }(z') ,
\end{align}
where $\phi_{\text{R},\lambda},\phi_{\text{L},\lambda}$ are respectively the right and left eigenfunctions of $ {\cal H}^\dagger - \mu D_\parallel(z)$.

At late times, the solution is dominated by the smallest eigenvalue  $\lambda_0(\mu)$ of the operator ${\cal H}^\dagger- \mu D_\parallel(z)$:
 \begin{equation}
g(\mu,t) = e^{-\lambda_0(\mu) t} \int_{-H}^H dz \ p_0(z)\phi_{\text{R},0 }(z) \ \int_{-H}^H dz'\   \phi_{\text{L},0 }(z') ,\label{Eq_Sol_q}
\end{equation}
where  
\begin{equation}
{\cal H}^\dagger\phi_{\text{R},0}(z) - \mu D_\parallel(z) \phi_{\text{R},0}(z)= \lambda_0(\mu)\phi_{\text{R},0}(z).
\end{equation}
Note that the prefactor is obtained from the initial condition $g(\mu,t=0)=1$. 
 Comparing with the definition  (\ref{843112}) of the cumulants via the generating function, this result then implies that at late times:
\begin{equation}
-\lambda_0(\mu) t= \sum_{n=1}^\infty \frac{ \mu^n}{n!}\langle Y_t^n\rangle_{\text{c}}.
\end{equation}
This formula means that all cumulants of $Y_t$ (and also all cumulants of $X_t$) scale as $t$ for large times:
\begin{align}
\langle Y_t^n\rangle_\text{c} \underset{t\to\infty}{\simeq}  u_n t , 
\end{align}
and the coefficients $u_n$ are found by considering the series expansion of $\lambda_0(\mu)$ near $\mu=0$, hence 
$u_n=-(\partial_\mu^{n}\lambda_0)_{\mu=0}$.  This gives an alternative method of computing the late time behavior of the cumulants using perturbation theory and one can check that it agrees with the Kubo formula used in the Letter for the second and fourth cumulants.

\subsection{Convergence to Gaussian statistics in the diffusive scaling regime.}
The goal of this section is to establish how the PDF $p(x,t)$ converges to a Gaussian when $x= \xi \sqrt{t}$ (this, in the diffusive regime), in the large time limit. We notice that  the   Fourier  transform of $p(x,t)$ written $\hat{p}(k,t)$ is obtained by setting $q=-i k$ in the moment generating function $\left[g(q,t)=\hat{p}(k,t)\right]$. Thus,  $p$ can be recovered by taking the inverse Fourier transform of $g$  :
\begin{align}
p\left(x=\xi \sqrt{t},t\right)  = \frac{1}{2\pi} \int_{-\infty}^\infty dk\   e^{i k \xi \sqrt{t} } \hat{p}(k,t) = \frac{1}{2\pi \sqrt{t}} \int_{-\infty}^\infty d\tilde{k} \  e^{i \tilde{k} \xi   }  \ g\left(-i \frac{\tilde{k}}{\sqrt{t}} ,t\right),
\end{align}
where we have set $\tilde{k}=k\sqrt{t}$ in the second equality. Now, we may formally use Eq.(\ref{843112}) to write:
\begin{align}
p\left(x=\xi \sqrt{t},t\right)   &= \frac{1}{2\pi \sqrt{t}} \int_{-\infty}^\infty d\tilde{k} \exp\left[i \tilde{k} \xi  + \sum_{n=1}^\infty  (-\tilde{k}^2)^n \frac{\langle Y_t^n\rangle_\text{c}}{t^n n!}\right] \\&
\underset{t\to\infty}{\simeq}  \frac{1}{2\pi \sqrt{t}} \int_{-\infty}^\infty d\tilde{k} \exp \left[
i \tilde{k} \xi  + \sum_{n=1}^\infty  (-\tilde{k}^2)^n \frac{u_n t }{t^n n!} \right] , 
\end{align}
where we have used the previously determined behavior $\langle Y^n\rangle_c\simeq u_n t$ at large times. We see that all terms with $n\ge 2$ in this expansion are proportional to $t^{1-n}$ and  can thus be treated as perturbative terms when $t\to\infty$. In particular, keeping only the term $n=2$ leads to:
\begin{align}
p\left(x=\xi \sqrt{t},t\right)    \underset{t\to\infty}{\simeq}  \frac{1}{2\pi \sqrt{t}} \int_{-\infty}^\infty d\tilde{k} \ e^{i \tilde{k} \xi  - u_1 \tilde{k}^2 } \left(1+\frac{u_2 \tilde{k}^4 }{2t}\right). 
\end{align}
Using $u_1= \langle D_{||}\rangle_0 $ and $u_2 = 2D_4$ and performing the integral leads to:
\begin{equation}
p\left(x=\xi \sqrt{t},t\right) \underset{t\to\infty}{\simeq}  \frac{1}{\sqrt{4\pi t \langle D_{||}\rangle_0 }}\exp\left(-\frac{\xi^2}{4 \langle D_{||}\rangle_0 }\right)\left\{1+ \frac{   D_4}{16\  t  \langle D_{||} \rangle_0^4}\left[12  \langle D_{||}\rangle_0^2(1-\xi^2)  + \xi^4\right]\right\}. \label{gaslt}
\end{equation}
The above derivation basically recovers the first terms of  the Gram Charlier series of type A used in the analysis of non-Gaussian \cite{cramerBook} random variables. We thus see that in the diffusive scaling regime, non-Gaussian statistics decay with time. This is of course compatible with the observation, made in the main text, that the non-Gaussianity parameter $\alpha(t)= \langle X^4_t\rangle_\text{c}/\langle X^2_t\rangle^2_\text{c}\sim 1/t$ at late times. In the next section, we will see  that the only trace of non-Gaussian statistics  at very late times is in the extreme value statistics where the central limit theorem does not apply.

\subsection{Gaussian tail of the displacement PDF at short times}

At short times, the displacement PDF reads:\cite{chubynsky2014diffusing,han2006brownian,chechkin2017brownian,lavaud2021stochastic}
\begin{align}
p(x,t)=\int_{-H}^H dz \frac{1}{\mathcal{Z}} e^{-\beta V(z)} \times \frac{e^{-\frac{x^2}{4D_\parallel(z)t}}}{\sqrt{4\pi D_\parallel(z)t}} \label{94021} ,
\end{align}
where $\mathcal{Z}=\int_{-H}^H  dz e^{-\beta V(z)}$. In this section, we analyze the tails of $p(x,t)$ given by the above expression for the parameters of the experiment. To make analytical calculations easier, we  use a simplified form for $D_\parallel(z)$ which is correct both near the wall (at leading order) and far from the wall (at next-to-leading order when $z\to \infty$)
\begin{align}
&D_\parallel(z)\simeq D_0\frac{\nu+2\tilde{z}}{2\tilde{z}+3\nu}, &\nu=\frac{9 a}{16}, \label{SimplProfile}
\end{align}
with $\tilde{z}=z+H$, so that $\tilde{z}=0$ corresponds to the situation where the particle touches the wall. This simplified expression differs by less than $8\%$ from Eq.~(17) in the main text. 
 Next, we note that in our experiment the length $l_{\textrm{B}}$ is much smaller than the distance between the plates of the channel, so that we can ignore the presence of the upper- wall. Most of the contribution to $p$ for large $x$ comes from the regions where the local diffusivity is largest, hence for large $z$, where $\beta V \simeq \tilde{z}/l_{\textrm{B}} $. 

We now write $p$ as
\begin{align}
&p(x,t)=\int_0^\infty d\tilde{z} \frac{1}{\mathcal{Z}}  \frac{e^{-f(\tilde{z}) }}{\sqrt{4\pi D_\parallel(z)t}}, &f(\tilde{z})=\frac{x^2}{4D_\parallel t}+ \frac{ \tilde{z}}{l_{\textrm{B}}}. \label{05421}
\end{align}
Let us define $z^*$ the value of $\tilde{z}$ for which $f'(\tilde{z})=0$. A simple calculation shows that
\begin{align}
&f(z^*)=\frac{x^2}{4D_0t}+x\sqrt{\frac{\nu}{l_{\textrm{B}}D_0t}}-\frac{\nu}{2l_{\textrm{B}}},\label{fStar}\\
&f''(z^*)=4\frac{\sqrt{D_0t}}{l_{\textrm{B}}^{3/2}\sqrt{\nu}x}, \ D_\parallel(z^*)=\frac{D_0 \sqrt{l_{\textrm{B}}} x}{2 \sqrt{D_0\nu t}+\sqrt{l_{\textrm{B}}} x}.
\end{align}
We now evaluate $p$ in Eq.~(\ref{05421}) with the saddle-point method: 
\begin{align}
p(x,t) 
\simeq \frac{1}{\mathcal{Z}}  \frac{1}{\sqrt{2 D_\parallel(z^*)t \ f''(z^*)}} e^{-f(z^*)  }\label{GaussianTailExpre}.
\end{align}
We note that $f(z^*)$, given by Eq.~(\ref{fStar}), is   quadratic in $x$, so that the tail of $p$ is clearly of Gaussian form for large $x$. This expression is in excellent agreement at large $x$ with the experimental and numerical results, see Fig.~2(f) in the main text.

\subsection{Extreme value statistics}
Here, we consider contributions to the PDF at large values of  $x$ corresponding to trajectories which diffuse much further than the typical (diffusive) ones. The probability of these rare events can be computed using large deviation theory. The reader is referred to \cite{touchette2009large} for a standard introduction written for physicists.
If we consider the moment generating function in the form of Eq. (\ref{gt1}), we see that for $q$ large, paths which are highly dispersed dominate the functional. The large deviation analysis for Taylor dispersion was carried out in Refs.~\cite{haynes2014dispersion,haynes2014dispersion1,Kahlen2017}  and we   adapt the analysis there to study the diffusing diffusivity model under consideration here. 

The tails of the displacement PDF are obtained by analyzing the limit $q\to\infty$ in the moment generating function. We are sampling the large dispersion regime where $Z_t$ stays close to $z^*$ where $D_{||}(z)$ attains its maximum. We make the physically relevant assumption that that $z^*$ is not located at the wall. We thus look for an eigenfunction $\phi_{\text{R},0}$ that is localized near $z=z^*$. 
Writing  $z=z^*+\frac{\zeta}{\vert q\vert ^\alpha}$, for some $\alpha>0$,  the eigenvalue equation for $\phi_{\text{R},0}(z)=\psi_0(\zeta)$ simplifies at leading order in $q\to\infty$: 
\begin{equation}
 \left\{\vert q\vert^{2\alpha} D_\perp(z^*) \frac{d^2}{d\zeta^2 }  + \lambda_0(q)+ q^2 \left[D_{||}(z^*) -\frac{1}{2}| D_{||}''(z^*)|\frac{\zeta^2}{\vert q\vert^{2\alpha}}\right]\right\}  \psi(\zeta)=0.
\end{equation}
We see that we have to take $\alpha=1/2$ so that   $\psi$  does not depend on $q$. We  recognize the quantum harmonic oscillator problem, so that we directly write the lowest eigenvalue solution:
\begin{align}
&\psi_0(\zeta)\propto \exp\left[-\sqrt{\frac{\vert D_{\parallel}''(z^*)|}{8 D_\perp(z^*)}}\zeta^2\right], & \lambda_0(q) = |q| \sqrt{\frac{|D_{||}''(z^*)|D_\perp(z^*)}{2}} -q^2D_{||}(z^*).  \label{EigsQuantum}
\end{align}
 
If we assume the large deviation form of $p(x,t)\sim e^{-t f(x/t)}$ (up to exponential prefactors), we find that
\begin{equation}
g(q,t) \sim \int dx \ e^{-t f\left(\frac{x}{t}\right)+qx } \sim \int d\xi \ e^{ t[q\xi -f(\xi)]} \sim \exp \left\{ \underset{\xi}{\max}[q\xi-f(\xi)]t \right\} ,
\end{equation}
where we have used the saddle point method and the notation $\xi = x/t$. Since  we already know that $g(q,t)\propto \exp[-t\lambda_0(q)]$, we see that
\begin{equation}
-\lambda_0(q)=\underset{\xi}{\max}[  q\xi- f(\xi)]. 
\end{equation}
Hence, the minimal eigenvalue $-\lambda_0(q)$ is the Legendre transform of the large deviation  function $f$  \cite{haynes2014dispersion,haynes2014dispersion1,Kahlen2017}. 
Inverting the Legendre transform we obtain:
\begin{equation}
f(\xi)= \underset{q}{\max} [q\xi+ \lambda_0(q)].
\end{equation}
The behavior of $f$ for large $x$ is obtained by taking the   Legendre transform of $\lambda_0(q)$ for large $\vert q\vert $ given by Eq.~(\ref{EigsQuantum}),  this leads to:
\begin{equation}
f(\xi) \underset{\vert\xi\vert\to\infty}{=} \frac{1}{4D_{||}(z^*)}\left[\xi +{\rm sign}(\xi)\sqrt{\frac{|D_{||}''(z^*)|D_\perp(z^*)}{2}}\ \right]^2 +\mathcal{O}(1),
\end{equation}
which is Eq. (\ref{main.f_late_time}) in the main text. This means that the PDF of the displacement $p(x,t)\sim e^{-t f(x/t)}$ has the form of a shifted Gaussian, with a diffusivity  given by the maximal value $D_{||}(z^*)$, however the weight of these paths is strongly suppressed by the way the Gaussian is centered.

Finally, the large deviation function for small $\xi$ can be simply computed from the second  moment and reads:
\begin{equation}
f(\xi) = \frac{\xi^2}{4\langle D_{||}\rangle_0},
\end{equation}
which matches with the result in the diffusive regime where $x=\mathcal{O}(\sqrt{t})$. 

\section{Experimental details}
The experimental data presented in the main text corresponds to spherical polystyrene colloids of nominal radius $1.5~\mathrm{\mu m}$ purchased from Polybead$^\copyright$. The ensemble of particles that are tracked have reached their equilibrium distribution due to the lag time between the insertion of the particles and the beginning of the measurement protocol. In the latter, a single sphere is three-dimensionally tracked using a self-calibrated interferometric method based on Mie Holography~\cite{lavaud2021stochastic}. 
\smallskip
\newline 
\indent A previously-calibrated plane wave (wavelength $532~\mathrm{nm}$) illuminates a dilute colloidal suspension. The light scattered by a given particle interferes with the incident beam in the focal plane of a $\times100$-objective and the interference pattern, called hologram, is magnified toward a CCD camera. Then, the strong dependencies of a hologram on both the physical properties and the position of the sphere lead to the precise measurement of the aforementioned characteristics.
The first 10000 holograms are fitted to determine the physical properties of the sphere, namely its radius and optical index. Those physical properties are then set and all the holograms are fitted, leading to the trajectory of the sphere, and, after its statistical analysis, to the observables depicted in the main text. 
\smallskip
\newline
\indent The experimental equilibrium PDF $p_0$ shown in~\cref{main.fig.exp}(a) of the main text is obtained by binning the $z$ position of the sphere on a logarithmic normal grid.
\smallskip
\newline
\indent The ensemble averages required in the computation of the experimental second and fourth cumulants (see~\cref{main.avd,main.kcumulant4}) depicted in~\cref{main.fig.exp}(c,d) of the main text, are obtained through sliding temporal averages, assuming ergodicity: 
\begin{equation}
    \langle X_t^n \rangle = \dfrac{1}{N} \, \sum_{j=0}^{N-1} \, \left[ x(t+t_j) - x(t_j) \right]^n \equiv \dfrac{1}{N} \, \sum_{j=0}^{N-1} \, \left[\Delta x (t \, \vert \, t_j)\right]^n\ ,
    \label{eq:slidingaverage}
\end{equation}
where $n, N \in \mathbb{N}$, $x(t)$ is the $x$ position of the sphere at time $t$, and where $t_j = j \, f_\mathrm{a}^{-1}$ with the frame rate $f_\mathrm{a} = 100~\mathrm{Hz}$ of the acquisition. 
\smallskip
\newline

\indent The experimental local diffusion coefficients depicted in~\cref{main.fig.exp}(b) of the main text are obtained by a stochastic force inference algorithm~\cite{frishman2020learning}.
An unbiased estimator $\hat{d}$ of the local diffusion coefficient in the $x$ direction (the adaptation to the $z$ direction being straightforward) is built as follows:
\begin{equation}
    \hat{d}(t_j) = \frac{\left[ \Delta x(\delta t \, \vert \, t_{j-1}) + \Delta x(\delta t \, \vert \, t_j) \right]^2}{4 \delta t} + \frac{\Delta x(\delta t \, \vert \, t_{j-1}) \Delta x(\delta t \, \vert \, t_j)}{2 \delta t}\ ,
    \label{eq:ronceray}
\end{equation}
where $\delta t$ is a chosen multiple of $f_a^{-1}$ and $\Delta x(\delta t \, \vert \, t_j)$ is the distance travelled by the sphere over a time $\delta t$ and starting at $t_j$.
Each of the above values of $\hat{d}$ corresponds to a given height $H+z$ and their distribution is estimated on a normal grid $\left\{ \Tilde{z} \right\}$ with a polynomial function basis of order 3: $\sum a_k(\Tilde{z})(H+z)^k$ , in which the $a_k$ are real functions. The accuracy of the method is confirmed \textit{a posteriori} by the agreement with the theoretical predictions, as shown in~\cref{main.fig.exp}.
We also note that the first term on the right-hand side of~\cref{eq:ronceray} arises from the temporal linearity of the MSD (see~\cref{main.avd} of the main text) while the second one is a correction that allows us to estimate accurately the local diffusion coefficients close to the surface ($H+z \le 100~\mathrm{nm}$) where the experimental data is scarce. 
\smallskip
\newline
\indent Finally, several observables that stem from the sphere's trajectory -- which include the ones described above -- depend on the physical parameters of the system, namely $B$, $l_\mathrm{D}$ and $l_\mathrm{B}$ defined in~\cref{main.Eq:PDF} of the main text. These parameters are thus fitted simultaneously to increase the method's precision.

\section{Numerical simulations}

We consider hereafter the three overdamped Langevin equations:
\begin{equation}
	\left\{
	\begin{aligned}
		\mathrm{d} X_t& = \sqrt{2 D_{\parallel}(Z_t)}\, dB_{x,t}  \\
		\mathrm{d} Y_t & = \sqrt{2 D_{\parallel}(Z_t)}\, dB_{y,t}  \\
		\mathrm{d} Z_t & = D_{\perp}'(Z_t)dt - \beta D_\perp(Z_t) V'(Z_t) dt+ \sqrt{2 D_{\perp}(Z_t)}\, dB_{z,t}\  ,
	\end{aligned}\right.
\label{eq:langevinMath}
\end{equation}
where the first right-hand-side term of the last equation corresponds to the spurious force in the Ito convention, and where $dB_{x,t}$, $dB_{y,t}$ and $dB_{z,t}$ are independent Brownian increments.
The simulation takes into account the bottom and top walls positioned at $\pm H_\mathrm{p}$. The potential $V(z)$ is given by Eq. (\ref{main.Eq:PDF}) of the main text.

The effective viscosity perpendicular to a single wall is given by Eq. (\ref{main.eq:d_perp}) of the main text. In the case of the relatively wide channel studied experimentally, we can use the superposition approximation and estimate the excess drag forces with respect to the bulk ones as the sum of the corresponding contributions from the individual surfaces. The total effective viscosities are thus given by:
\begin{equation}
\mu_i^T(z)-\mu_0 \simeq \mu_i^{+}(z)-\mu_0 + \mu^{-}_i(z)-\mu_0\ ,
\end{equation}
with $i \in \{\|, \perp\}$, and where $\mu_i^{\pm}(z)$ denote the individual effective viscosities near both walls. The Stokes-Einstein relation then gives:
\begin{equation}
	D_i(z) \simeq \frac{ k_\mathrm{B}T}{6 \pi a \left[ \mu_i^{+}(z)+\mu^{-}_i(z) - \mu_0 \right] }\ .
	\label{eq:langevinDescretised}
\end{equation}
We plot $D_{||}(z)$ and $D_\perp(z)$ for $H_\text{p}=40\mathrm{\mu m}$ in Fig. \ref{figS1}(a), and for $H_\text{p}=5.5\mathrm{\mu m}$ in Fig. \ref{figS1}(b). Also shown for comparison are the parabolic approximations for the narrow-channel case.

\begin{figure}[th!]
	\includegraphics{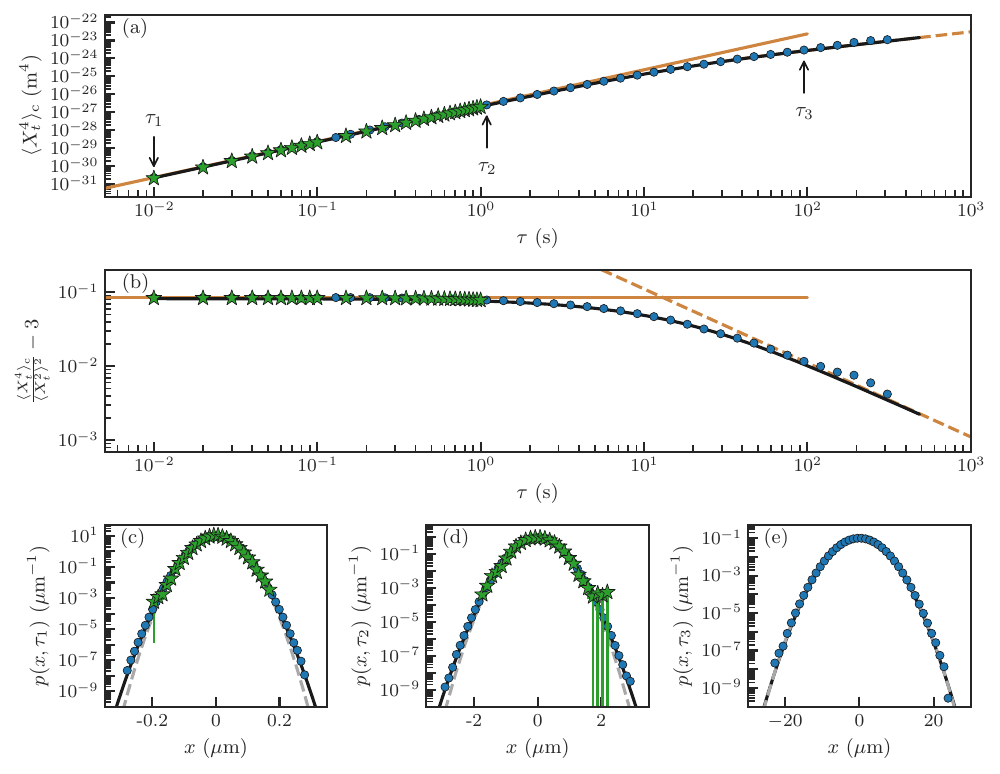}
	\caption{Fourth cumulant (a), and rescaled fourth cumulant (b) as functions of time, for $H_{\textrm{p}}=40\,\mu$m. Green stars: experimental data. Blue dots: simulation data. Orange solid and dashed lines: asymptotic predictions (see main text). Solid black lines: exact theory at all times (see main text). The three last panels correspond to the PDF $p(x,t)$ for (c) $t=\tau_1=0.01$s, (d) $t=\tau_2=1.09$s and (e) $t=\tau_3=95.4$ s. Green stars: experimental data. Blue dots: simulation data. Grey dashed lines: late time Gaussian distribution given by the first term of Eq.~(\ref{gaslt}). Solid black lines: exact prediction obtained by numerically inverting the Fourier transform $\hat{p}(k,t)=g(q=-ik,t)$ given by Eq.~(\ref{Eq_Sol_q}).}
	\label{figS2}
\end{figure}

\begin{figure}[th!]
	\includegraphics{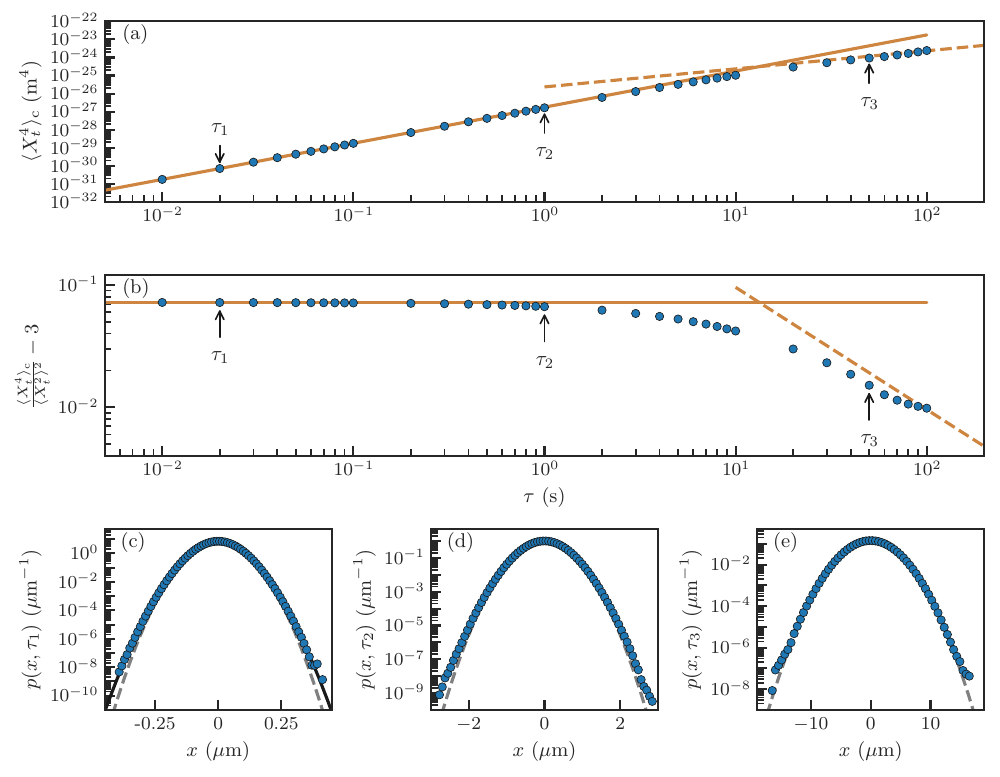}
	\caption{Fourth cumulant (a), and rescaled fourth cumulant (b) as functions of time, for $H_{\textrm{p}}=5.5\,\mu$m. Blue dots: simulation data. Orange solid and dashed lines: asymptotic predictions (see main text). Solid black lines: exact theory at all times (see main text). The three last panels correspond to the PDF $p(x,t)$ for (c) $t=\tau_1=0.01$s, (d) $t=\tau_2=1.09$s and (e) $t=\tau_3=95.4$ s. Blue circles: simulation data. Grey dashed lines: late time Gaussian distribution given by the first term of Eq. (\ref{gaslt}). Solid black lines: exact prediction obtained by numerically evaluating the expression of $p$ at short-times given by Eq.~(\ref{94021}). 
}
	\label{figS3}
\end{figure}

We discretize \cref{eq:langevinMath} by using an Euler scheme where solutions are approximated by $X_t(t) \approx X_n(t_n)$, $Y_t(t) \approx Y_n(t_n)$ and $Z_t(t) \approx Z_n(t_n)$, with  $t_n = n \Delta  t$, $\Delta t$ being the simulation time step. The increments  $dB_{k,t}$ ($k \in \{x, y, z\}$) are approximated by $\Delta B_{k,n}= W_{k,n}$, where  $W_{k_n}$ are independent Gaussian-distributed random variables of zero mean and unit variance. This leads to the discrete stochastic equations:
\begin{equation}
	\left\{
	\begin{aligned}
		X_{n+1} & = X_{n} + \sqrt{2 D_{\parallel}(Z_{n})}\, W_{x,n}\sqrt{\Delta t}  \\
		Y_{n+1} & = Y_{n} + \sqrt{2 D_{\parallel}(Z_n)}\, W_{y,n}{\sqrt{\Delta t}}    \\
		Z_{n+1} & = Z_{n} + D_{\perp}'(Z_n)\Delta t - \beta D_\perp(Z_n) V'(Z_n)\Delta t + \sqrt{2 D_{\perp}(Z_n)}\, W_{z,n}\sqrt{\Delta t}\   .
	\end{aligned}\right.
\label{langevinDescretised}
\end{equation} 

We numerically integrate (\ref{langevinDescretised}) with $\Delta t = 0.01~\mathrm{s}$, for a total time of $1000~\mathrm{s}$, with identical physical parameters as the experimental ones. The system is allowed to first equilibrate in the vertical direction. From approximately 12 million trajectories, we extract the numerical fourth cumulant $\langle X^4_t \rangle_\mathrm{c}$ at all times using the PDF $p(x, \tau)$ of displacements $x=X_{t+ \tau}-X_{t}$ generated from sliding temporal averaging. Specifically, the fourth cumulant is numerically calculated from:
\begin{equation}
	\langle X^4_{t= \Delta \tau} \rangle_\mathrm{c} = \int_{- \infty}^{+ \infty} u^4 P(u, \Delta \tau) \, \mathrm{d}u - 3 \left[\int_{- \infty}^{+ \infty} u^2 P(u, \Delta \tau) \mathrm{d}u \right]^2 .
\end{equation}

Finally, as shown in \cref{main.fig.exp}(d) of the main text and in Fig.~\ref{figS2}, the numerical results are in good agreement with both theoretical and experimental results, for a channel with $H_\text{p}= 40\, \mathrm{\mu m}$.
We have also carried out numerical simulations for a much narrower channel, \textit{i.e.} with $H_\text{p}= 5.5\, \mathrm{\mu m}$, but otherwise using the same parameters as the ones in the experimental setup. The results are shown in Fig.~\ref{figS3}. 

\bibliographystyle{naturemag}
\bibliography{Alexandre2022}